# Leveraging Machine Learning for Wi-Fi-based Environmental Continuous Two-Factor Authentication

Ali Abdullah S. AlQahtani, *Member, IEEE*, Thamraa Alshayeb, Mahmoud Nabil, and Ahmad Patooghy, *Senior Member, IEEE*

***Abstract*—The traditional two-factor authentication (2FA) methods primarily rely on the user manually entering a code or token during the authentication process. This can be burdensome and time-consuming, particularly for users who must be authenticated frequently. To tackle this challenge, we present a novel 2FA approach replacing the user's input with decisions made by Machine Learning (ML) that continuously verifies the user's identity with zero effort. Our system exploits unique environmental features associated with the user, such as beacon frame characteristics and Received Signal Strength Indicator (RSSI) values from Wi-Fi Access Points (APs). These features are gathered and analyzed in real-time by our ML algorithm to ascertain the user's identity. For enhanced security, our system mandates that the user's two devices (i.e., a login device and a mobile device) be situated within a predetermined proximity before granting access. This precaution ensures that unauthorized users cannot access sensitive information or systems, even with the correct login credentials. Through experimentation, we have demonstrated our system's effectiveness in determining the location of the user's devices based on beacon frame characteristics and RSSI values, achieving an accuracy of 92.4%. Additionally, we conducted comprehensive security analysis experiments to evaluate the proposed 2FA system's resilience against various cyberattacks. Our findings indicate that the system exhibits robustness and reliability in the face of these threats. The scalability, flexibility, and adaptability of our system render it a promising option for organizations and users seeking a secure and convenient authentication system.**

## I. Introduction

As technology continues to advance at a rapid pace, the need for robust security measures to safeguard sensitive information and data has become paramount. The Identity Theft Resource Center (ITRC) report reveals that data breaches have surged by over 68% in 2020, with a total of 1,862 incidents [1]. In response to these threats, the adoption of Two-Factor Authentication (2FA) has seen a significant growth. Google's 2020 study indicates that 150 million people employ 2FA to secure their accounts, the number is expected to rise as cybersecurity takes precedence among individuals and organizations [2]. Microsoft's research corroborates the effectiveness of 2FA, demonstrating that it can thwart 99.9% of automated cyberattacks [3]. As a result, 2FA has become an indispensable tool for protecting sensitive data. A survey by the Ponemon Institute further underscores its importance, showing that 56% of U.S. organizations have implemented 2FA for at least a portion of their workforce [4].

As 2FA necessitates two authentication factors from users to augment the security beyond a mere password, the adoption of the second factor has become a research question and investigated in [5]–[7]. Researchers have used a range of factors encompassing knowledge-based, possession-based, inherence-based, location-based, behavioral-based, and ambient-based factors [8]. Each approach aims to uniquely verify a user's identity while bolstering information security.

However, existing 2FA techniques possess limitations, including user inconvenience, susceptibility to human error, single-point authentication, restricted adaptability, scalability challenges, and inflexibility. To tackle these issues, this paper introduces a novel 2FA system that harnesses Wi-Fi radio waves and Machine Learning (ML) to authenticate a user's identity. The proposed system strives to deliver a seamless, user-friendly 2FA experience that minimizes the need for users to supply additional authentication factors beyond their primary login credentials. Furthermore, the proposed system addresses the limitations of existing 2FA methods, enhancing scalability, adaptability, and flexibility.

This paper is an extension of our previously published paper [9]. The new contributions with respect to the older version are listed below:

- We have incorporated four more ML models to develop an ML-based 2FA system that leverages Wi-Fi access point broadcast messages and RSSI values for reliable and secure user authentication. These models help eliminate the need for additional user input or action, reducing human error risks and improving the overall user experience.
- We have incorporated continuous authentication into our system, which enhances security by constantly verifying the user's identity while accessing protected resources, effectively thwarting unauthorized access.
- We conducted a Feature Importance analysis, which provides insights into the significance of each feature in the

A. AlQahtani (the corresponding author) and A. Patooghy are with the Department of Computer Systems Technology, North Carolina A&T State University, Greensboro, NC, USA, 27411.
E-mails: AlQahtani.aasa@gmail.com, apatooghy@ncat.edu
T. Alshayeb is with the Department of Physics & Astronomy, George Mason University, Fairfax, VA, USA, 22030.
E-mail: Alshayeb.t@gmail.com
M. Nabil is with the Department of Electrical & Computer Engineering, North Carolina A&T State University, Greensboro, NC, USA, 27411.
E-mail: mnmahmoud@ncat.edu

classifications made by the ML models in the proposed system.
- We evaluated the computation overhead of the proposed system in implementing the second layer of authentication using different ML models.
- We have conducted a comprehensive security analysis of our proposed system to evaluate its resilience against potential cyberattacks, including evasion attacks, model extraction attacks, and radio frequency signal interference attacks.

The remainder of this paper structure is organized as follows: Section II discusses ML-based secure authentication and continuous 0E2FA methods for various applications. Section III presents a detailed description of the network and threat models for the proposed system. In Section IV, we present the proposed system and how it uses beacon frames and RSSI values to verify a user's identity. Section V describes the experiments conducted to evaluate the proposed system, including the dataset used, the performance metrics calculated, and the results obtained. In section VI, we discuss the security analysis of the proposed 2FA system by examining its vulnerability to various cyberattacks and evaluating its resilience against them. Section VII discusses the proposed system's features, including reliability, zero-effort authentication, continuous authentication, adjustability, scalability, flexibility, and a one-time login solution. Section VIII concludes and summarizes the key points and contributions of the proposed system.

## II. Related Work

Various secure authentication mechanisms utilizing ML have been proposed for different applications [10]–[14]. For instance, Gupta et al. [15] proposed a secure authentication mechanism that uses ML and nonce-based systems for a telecare medical information system. Punithavathi et al. [16] introduced a cloud-based cancelable biometric authentication system for IoT devices.

In [17], the authors propose a deep-learning-based active authentication method that utilizes sensors in consumer-grade smartphones to authenticate users. Furthermore, a mouse data protection technology is introduced in [18] that generates random mouse positions to protect mouse data and uses ML to verify its security. A novel technique for secure access to smartphones utilizing piezoelectric touch sensing supported keystroke dynamics to authenticate users was proposed [19]. An IoT authentication system based on ML that uses human impedance as a user identifier is also introduced in [20]. Furthermore, the integration of ML techniques into a DevOps ecosystem to develop a Risk Authentication/Assessment Decision Engine (RADE) that estimates the risk level of each authentication attempt is presented in paper [21]. Lastly, a new two-factor authentication scheme based on real-time keystroke dynamics using the K-nearest neighbor classification algorithm is introduced in [22].

In the field of secure communication, two recent papers propose novel methods for authentication using physical layer features and deep learning algorithms. The first paper presents a framework for preventing spoofing attacks in Controller Area Networks (CANs) by exploiting physical layer features and using reinforcement learning to choose the authentication mode and parameter [23]. The proposed scheme employs a deep learning approach to further enhance authentication efficiency.

In [24], Gabriele *et al.* investigate the use of AI-based solutions for physical-layer authentication of Low-Earth Orbit (LEO) satellites, a challenging scenario due to non-standard electronics and unique attenuation and fading characteristics. The study uses Convolutional Neural Networks (CNN) and autoencoders to authenticate satellite transducers with high accuracy but highlights the potential limitations due to the high number of I-Q samples required and the low bandwidth of satellite links.

Various Zero-Effort Two-Factor Authentication (0E2FA) mechanisms that aim to authenticate users without requiring extra user interaction were published [25]–[29]. The paper, [30], leverages environmental Bluetooth Low Energy (BLE) signal characteristics for co-location detection, while SoundAuth [31] uses ambient audio signals and ML techniques for authentication. The proposed system in [32] captures gait patterns using a smartphone and smartwatch/bracelet, and Vibe [33] uses vibration communication for user authentication.

Another study [34] presents a smart hotel access system that uses Near Field Communication (NFC) Host-Card Emulation application for authentication. These mechanisms provide more secure and reliable solutions for zero-effort authentication while maintaining user-friendliness. They have been tested and evaluated for usability and security, and they demonstrate good resistance to attacks and high detection accuracy. The vulnerability of touch-based continuous authentication systems (TCAS) to active adversarial attacks is investigated in [35].

The significant role of security incidents caused by insiders in attacks against organizations is addressed in [36], which proposes a continuous authentication solution for large multi-site enterprises based on Apache Spark, Apache Cassandra, Kafka, and a MySQL database.

A novel authentication mechanism for smartphones that utilizes gait patterns and keystroke dynamics as behavioral biometrics to establish a multimodal biometrics profile is proposed in [37]. The experimental results demonstrate the robustness and security of the proposed method against different types of attacks.

Finally, a behavioral biometrics approach for exoskeletons using wearable sensors to ensure the user's identity and authority during operation is proposed in [38], which could be useful for other wearables used in robot control.

Recent advancements in adversarial attack methodologies proposed for DL-based wireless signal classifiers aimed at anti-eavesdropping have been highlighted in [39]. In contrast, our research primarily centers on developing two-factor authentication. By utilizing ML for continuous identity verification based on environmental features, we emphasize providing secure and convenient user authentication without directly addressing adversarial attack techniques.

## III. Network & Threat Models

This section provides a detailed description of the network model used in the proposed 2FA system, as well as explores the threat model of the system.

### A. Network Model

In the proposed 2FA system, at least one Wi-Fi Access Point (AP) must be present in the user's environment. This AP will transmit beacon frames containing unique characteristics. For system use, the user needs two Wi-Fi-enabled devices (e.g., a smartphone and a laptop). The laptop/PC acts as a login device, and the smartphone serves as a mobile device. Both devices must contain applications that collect and transmit data to the authentication entity. Below is a list of the required components, see Figure 1:

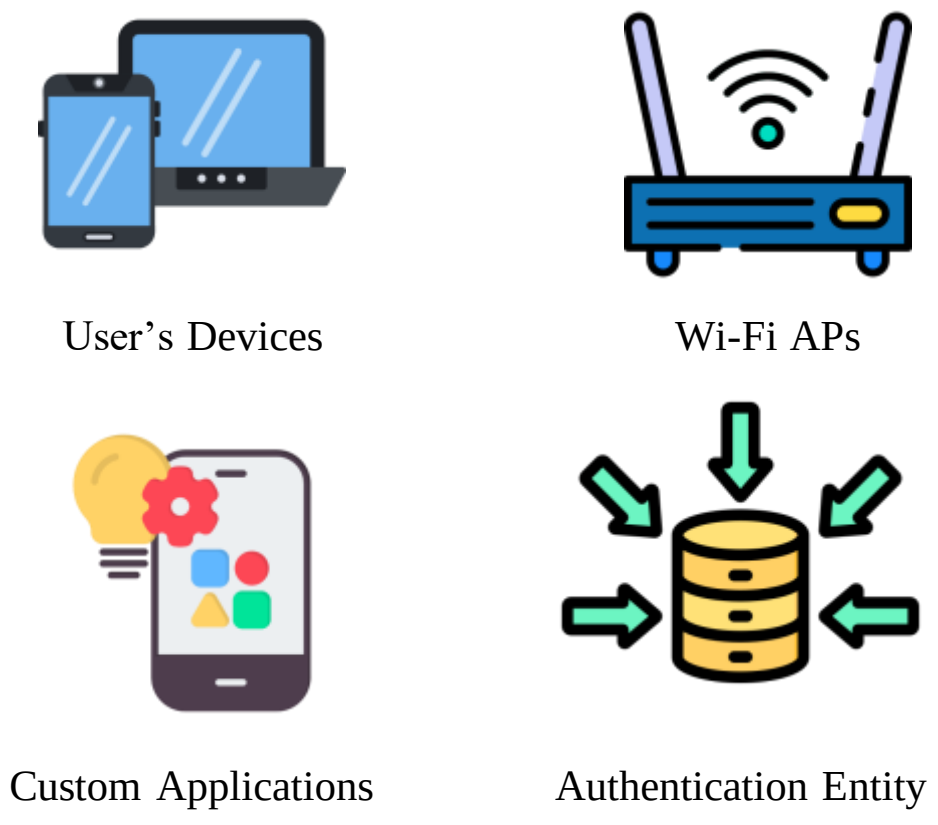

Fig. 1: Required Components; icons from [40]

1) Device Requirements: the proposed system necessitates the utilization of two distinct devices for authentication purposes: a login device and a mobile device. The login device, such as a computer, allows the user to access the system, while the mobile device, typically a smartphone, is carried by the user to send and receive beacon frames.
2) APs: The system relies on APs to disseminate beacon frames for user authentication. These APs can be any wireless AP compatible with IEEE 802.11 standards, including Wi-Fi APs.
3) Custom Applications: To facilitate the authentication process, our system mandates the use of specialized applications on both the login and mobile devices. These applications are responsible for the transmission and reception of beacon frames and for communication with the server and database.
4) Authentication Entity: This component is in charge of storing user authentication data and orchestrating the authentication procedure. By receiving and processing beacon frames and RSSI values from user devices (i.e., a login device and a mobile device), it makes informed authentication decisions based on the properties of these frames. The authentication entity consists of the following sub-components.
   a) A database that stores the selected beacon frame characteristics i.e., SSID, BSSID (i.e., uses only in step 4 in Figure 2), and Wi-Fi radio waves frequency) and RSSI values collected from the user's devices, as well as the user's login credentials and other necessary information required for the authentication process. In general, Wi-Fi beacon frames contain various characteristics that serve specific purposes in network operations. However, not all of these characteristics have a direct impact on device localization. By focusing on the ones that directly affect device localization (i.e., SSID, BSSID, Wi-Fi radio wave frequency) and RSSI - we can streamline the localization process, reduce computational complexity, and develop an effective solution for real-world applications.
   b) An authentication module that utilizes an ML algorithm analyzes the collected data from the user's devices and makes the final decision on whether to grant the user access to the authentication entity.

During the registration phase, the users must activate and create their account through the custom mobile application. Once logged in, the application runs in the background, waiting for instructions from the authentication entity. Every time the user wants to use the system, the application automatically scans their environment and transmits the collected data to the authentication entity. The process is seamless and does not require the user to take any additional actions through the mobile application.

### B. Threat Model

The proposed system considers two primary categories of potential attackers: internal and external. Individuals who possess authorized access to the system, such as employees, contractors, or those with insider knowledge of the system's architecture, algorithms, and data, fall under the internal attackers category. Such attackers may exploit their access or knowledge to compromise the system's security from within by manipulating input data or ML models to achieve their objectives.

External attackers, on the other hand, are individuals or entities without authorized access to the system. These attackers typically exploit vulnerabilities in the system from outside by intercepting communications, finding weaknesses in the algorithms, or compromising the infrastructure that supports the system.

The proposed 2FA system, relying on radio waves and ML, is susceptible to a range of security threats that can undermine its performance and security. Attackers can exploit weaknesses in the ML models, manipulate input data, or obstruct communication channels, leading to the system's compromise. The considered threats in this work include *evasion*, wherein malicious inputs are created to bypass the detection mechanisms, *model extraction*, which involves reverse-engineering the ML model to gain unauthorized access, and radio frequency signal interference attacks that disrupt the communication channels by emitting conflicting signals, reducing the system's ability to operate effectively. These threats are discussed and examined in Section VI.

## IV. The Proposed System

This section provides a detailed description of the proposed system. Specifically, we will discuss the authentication process and the features of the proposed system. Before utilizing the system, users must create a profile and download the required applications; discussed early in Subsection III-A.

### A. Authentication Phase

The proposed authentication follows a multi-layered approach to access protected entities. The first layer requires users to enter a valid username and password for authentication. After this, the second authentication layer requires users to satisfy two criteria. The first criterion mandates that both the user's devices (i.e., a login device and a mobile device) are present in a predefined number of overlapped Wi-Fi access points that are visible to both devices. The second criterion demands that the devices are within a specific proximity threshold; more details will be presented in Subsection IV-B3.

We assume that the determination of the number of overlapping Wi-Fi APs and the proximity threshold is conducted through an administrative process that considers the particular security needs of the protected entity.

The proposed authentication process aims to balance convenience and security by minimizing the user's burden while ensuring reliable authentication. To provide readers with a clearer understanding of the system configuration and underlying processes, we further detail our algorithm and models. The subsequent sections will delve into the specifics. Figure 2 illustrates the system configuration and steps involved in the process:

1) The user attempts to access the authentication entity and enters his/her login credentials (username and password).
2) The server validates the credentials by checking them against the stored user profile.
3) The server triggers the user's devices (i.e., a login device and a mobile device) to collect the selected beacon frame characteristics and measure RSSI values from the Wi-Fi APs in the user's environment, and transmit them to the server.
4) In order to confirm the location of the user's devices, the server checks that both devices can detect a predefined number of overlapping Wi-Fi APs. This is done by comparing the unique identifiers (SSIDs and BSSIDs) of the Wi-Fi APs. Both devices will scan and detect Wi-Fi APs in his/her vicinity and create a list of the Wi-Fi APs that have matching SSIDs and BSSIDs. By identifying overlapping Wi-Fi APs, the server can confirm that the user's devices are in close proximity to one another.
5) The server uses ML to analyze the collected data from the user's devices (i.e., a login device and a mobile device) to determine if the devices are within the predefined threshold or not.
6) The final decision is made based on the joint success of steps 4 and 5. If both are successful, access is granted; if not, access is denied.

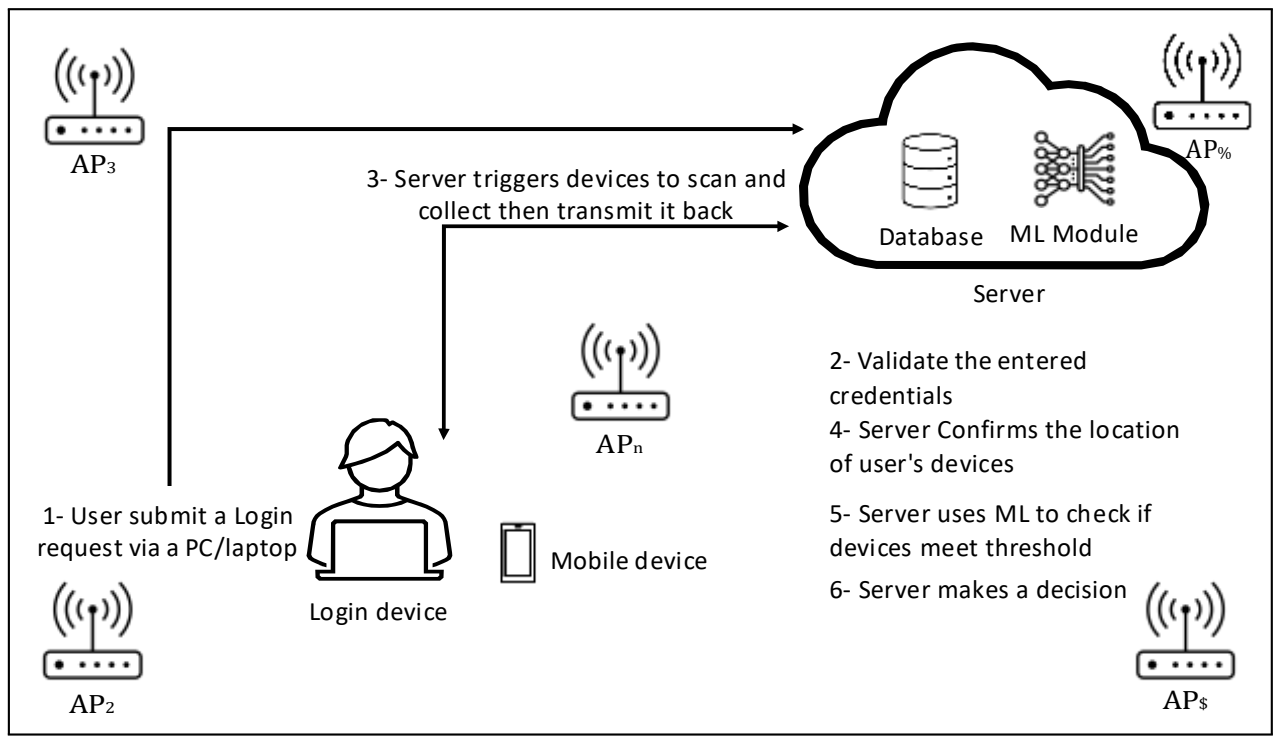

Fig. 2: System Configuration & Authentication Process

It is essential to note that while our authentication may seem to follow a binary decision framework (access granted or denied), the underlying processes involve intricate analyses. The binary classification approach was chosen for its suitability in authentication scenarios. Here, a user is either authenticated or not, making it inherently binary. However, the complexity arises in making this binary decision, which involves multi-layered checks, continuous monitoring, and ML-based evaluations. These intricacies go beyond mere binary results and are pivotal for ensuring robust and reliable authentication.

### B. The Proposed System Features

In this subsection, we will explain the features of our system incentives. Then, we will compare these features against the discussed related works in Subsection IV-C.

*1) Zero-effort:* The proposed system has a "zero-effort" feature or "Noninteractive", which significantly enhances the user experience by automating the authentication process, eliminating the need for additional input or action. By employing ML to analyze the selected beacon frame characteristics and RSSI values from users' devices (i.e., a login device and a mobile device) the system efficiently implements the second layer of authentication and verifies their identity; steps 4 and 5. This seamless approach not only boosts user satisfaction but also bolsters security through the reduction of human error risks, ultimately resulting in a streamlined and secure authentication experience.

*2) Continuous Authentication:* The continuous authentication feature in the proposed system enhances security by constantly verifying a user's identity while they access protected resources; repeating step 5 in Figure 2. Utilizing ML technology, the system analyzes distinctive environmental characteristics and signals to ensure that the user's devices remain co-located. If the devices are no longer in the same location, the session is promptly terminated. This approach not only improves security by thwarting unauthorized access but also offers user convenience and alleviates the need to remember to log out, resulting in a seamless and secure experience; this feature will be examined in the experiment presented in Section V-D and achieved a 100% success rate in terminating the session when the user's devices were not co-located.

*3) Adjustability:* The adjustability of the proposed system plays a crucial role in providing a high level of security, as it facilitates the customization of the access policy process according to the distinct needs of each protected entity. By modifying key factors, such as the number of Wi-Fi APs involved in users' identity verification and the distance threshold (i.e., used in the authentication process, steps number 4 and 5 in Figure 2), the system can be fine-tuned to address the security requirements of a wide range of systems and data types. This adaptability enables a tailored security solution that effectively accommodates varying levels of protection, ensuring that each entity's unique demands are met without compromising the overall performance and reliability of the system.

*4) Flexibility:* The proposed system's use of broadcast messages is a crucial element of its flexibility. By relying on broadcast messages (i.e., used in the authentication process, step number 3), the system can easily adapt to different settings without significant modifications. This allows the system to be deployed in various environments without requiring extensive customization, reducing the time and cost associated with implementation. The use of ML further enhances the system's flexibility by enabling it to adjust to the unique characteristics of each user's environment. This means that the system can accurately identify users, regardless of their location or device. As a result, the proposed system is a highly versatile solution that can be applied to a wide range of environments, providing an adaptable and robust security solution.

### C. Comparison with the Discussed Related Works

Although several 2FA methods have been proposed and disucsieed in Section II, we believe that the community needs to put some efforts to progress the field. Reviewing the related works show that almost each and every of the proposed methods suffer from at least one of the desired criteria for 2FA e.g., interactability, continuum, adjustability, and flexibility. Table I summarizes this claim and compares our proposed method against methods of the literature. Later on, we will discuss how each of these criterion is supported in our method in their respective subsections.

TABLE I: Comparison with The Discussed Works, Where ✓: Feature Supported, and ✗: Not Supported

| Study | Noninteractive | Continuum | Adjustability | Flexibility |
|---|---|---|---|---|
| [10]–[24] | ✗ | ✗ | ✗ | ✗ |
| [25]–[38] | ✓ | ✓ | ✗ | ✗ |
| Ours | ✓ | ✓ | ✓ | ✓ |

## V. Experiment

In this section, we describe the experimental setup utilized to evaluate the proposed authentication system.

### A. System Configuration

The system configuration consisted of the following components.

1) Wi-Fi APs: A total of ten APs were detected by the system utilizing publicly available Wi-Fi APs in a North Carolina A&T research laboratory on a typical school day to simulate a busy building setting. It is noteworthy that the detection of radio waves from the APs did not require any connection to them.
2) User devices: Raspberry Pi 3 Model B boards were employed to mimic the users' two devices, namely, the mobile device and the login device.
3) Server: A desktop computer with Ubuntu 16.04 LTS (64-bit) as the operating system was utilized to run the computer and host the ML module. Additionally, PhPmyadmin was used to create and execute the database.

### B. Data Collocation Phase

The purpose of the data collocation phase was to gather sufficient data to train the proposed system's ML module to determine the location of the user's devices based on the chosen beacon frame characteristics and RSSI values from Wi-Fi APs. To collect the data, a maximum threshold distance of 7 feet between the user's devices was established.

Two datasets were then collected: an *"authentic"* dataset with data collected when the two Raspberry Pis were within 7 feet of each other, and an *"unauthorized"* dataset with data collected when the distance between the Pis was greater than 7 feet (a minimum distance of 7.5 feet). To diversify the data samples, the Pis were repeatedly placed at varying distances for both datasets.

This approach helped identify the *"gray area"* between the acceptable threshold distance and the distance at which access should be denied. A total of 4,825 data samples were collected from two Raspberry Pis for the two datasets, with 2,442 samples in the *"authentic"* dataset and 2,383 samples in the *"unauthorized"* dataset. This near-equal distribution helps in preventing biases in the ML model towards either category. These samples were collected from 10 different Wi-Fi access points located in various positions within the experimental areas and at different times.

The resulting datasets consisted of six columns: "RPi", "SSID,", "Frequency (Hz)", "RSSI (dBm)", "Location", and "Label." In our datasets, the independent variables include "RPi" (which Raspberry Pi collected the data), "SSID" (name of the Wi-Fi AP), "Frequency (Hz)" (frequency of the Wi-Fi AP), and "RSSI (dBm)" (RSSI value in dBm). These variables represent the factors that we manipulated or observed to see their effect on the dependent variable. The dependent variable in our study is "Label," a binary variable indicating whether the sample is "authentic" (1) or "unauthorized" (0). This dependent variable reflects the outcome we are interested in predicting based on the independent variables.

The datasets were curated to ensure a balance between the "authentic" and "unauthorized" samples.

For empirical evaluation, we employed the N-fold cross-validation technique. Specifically, we utilized 5-fold cross-validation in our analysis. This involved dividing the dataset into five parts, using four parts for training and one part for testing iteratively.

| \ | RPi | SSID | Frequency (Hz) | RSSI (dBm) | Location | Label |
|---|---|---|---|---|---|---|
| 0 | 1 | 1 | 2412 | -65 | 1 | 1 |
| 1 | 1 | 2 | 2437 | -68 | 1 | 1 |
| 2 | 1 | 3 | 2462 | -75 | 1 | 1 |
| 3 | 1 | 4 | 2412 | -82 | 2 | 1 |
| 4 | 1 | 5 | 2437 | -79 | 2 | 1 |

Fig. 3: Dataset Visualization

In the evaluation phase, we partitioned our labeled data into a training set (80%) and a testing set (20%). Six different ML models were employed to evaluate the proposed system, namely Decision Tree (DT), K-Nearest Neighbors (K-NN), Random Forest (RF), Support Vector Machine (SVM), Naive Bayes (NB), and Logistic Regression (LR). The selection of these models was based on their effectiveness in similar classification tasks. We computed the confusion matrix for each model and used the five standard evaluation metrics for classification tasks in ML: accuracy, sensitivity, specificity, F1 Score, and precision. The confusion matrices for the DT, KNN, RF, SVM, NB, and LR models are presented in Figure 4.

TABLE II: Evaluation Results

| Model | Accuracy | Sensitivity | Specificity | F1 Score | Precision |
|---|---|---|---|---|---|
| DT | 0.924 | 0.918 | 0.932 | 0.926 | 0.934 |
| KNN | 0.923 | 0.911 | 0.936 | 0.924 | 0.938 |
| RF | 0.922 | 0.92 | 0.925 | 0.924 | 0.929 |
| SVM | 0.903 | 0.895 | 0.91 | 0.904 | 0.914 |
| NB | 0.885 | 0.849 | 0.923 | 0.884 | 0.921 |
| LR | 0.881 | 0.891 | 0.87 | 0.885 | 0.879 |

As can be seen from Table II, the DT and KNN models perform the best overall, with DT achieving the highest accuracy and F1 Score, and KNN obtaining the highest specificity and precision. RF also performs well with the highest sensitivity.

### C. Feature Importance

The Feature Importance analysis was conducted to understand the significance of each feature in the classifications made by the ML models within the proposed system. This information is instrumental in identifying the key features that drive the model's accuracy. We determined the feature importance for DT, RF, KNN, SVM, and LR. However, this metric is not applicable to the NB model. Since NB is a probabilistic model, it assumes conditional independence among features given the target class and does not account for feature interactions or correlations when classifying.

Identifying feature importance enables us to pinpoint the most crucial factors in generating accurate predictions. In our Python implementation, we used the *feature_importances_* attribute to directly obtain the importance of each feature for DT and RF models. For KNN, we employed the *mutual_info_classif* function from the *sklearn.feature_selection* module to compute the mutual information between each feature and the target variable. Lastly, we determined feature importance for SVM and LR models by utilizing the absolute value of the coefficients, accessed through the *coef_* attribute.

Table III presents our results. In the DT model, the most important feature is RSSI (dBm), with a 70.79% score, followed by Frequency (Hz) at 17.42%. For the KNN model, RSSI (dBm) remains the most crucial feature, contributing 30.10%, while Frequency (Hz) accounts for 19.75%. In the RF model, RSSI (dBm) again holds the highest importance at 54.56%, with Frequency (Hz) coming in second at 27.70%. Intriguingly, the SVM model identifies SSID as the most important feature with a 65.20% score, and RSSI (dBm) follows at 21.75%.

TABLE III: Feature Importance Results

| Model | RPi | SSID | Frequency (Hz) | RSSI (dBm) | Location |
|---|---|---|---|---|---|
| DT | 0.18% | 7.82% | 17.42% | 70.79% | 3.79% |
| KNN | 0.34% | 0.18% | 19.75% | 30.10% | 4.80% |
| RF | 1.13% | 7.28% | 27.70% | 54.56% | 9.32% |
| SVM | 0.01% | 65.20% | 13.04% | 21.75% | 0.01% |
| LR | 0.61% | 55.20% | 1.19% | 37.09% | 5.91% |

From Table III, it is evident that RSSI (dBm) is the most significant feature for DT, KNN, and RF models, while the SVM and LR model prioritizes SSID. Other features, such as RPi, Frequency (Hz), and Location, exhibit varying levels of importance across different models.

### D. Continuous Authentication

To simulate continuous authentication using the proposed system, an algorithm was developed and translated to Python code that both mimics the system and continuously monitors the user's environment. The continuous authentication process, outlined in Figure 5, involves monitoring the location of the user's devices (i.e., a login device and a mobile device) and terminating the session if they are no longer in the same location.

The continuous authentication feature collects new Wi-Fi characteristics data from the user's APs using a loop, with each iteration processing the collected data using previously trained ML models, including DT, KNN, RF, SVM, NB, and LR. The loop identifies the label for the new data using each model in the list. If any of the models classifies a negative label, indicating that the user's devices (i.e., a login device and a mobile device) are no longer co-located, the session terminates, and the loop exits. If the label is positive, the loop waits for a specified interval of 30 seconds, which is most commonly supported by devices for scanning at intervals of 30 seconds or longer, before conducting another check.

This process repeats continuously until the session is terminated by a negative label classification or a user ends the session. In the experiment, the system was tested by increasing the distance between two Raspberry Pis beyond the threshold of 7 feet after access was granted. Once the proposed system detected that the user's devices (i.e., a login device and a mobile device) were no longer in range (after the predetermined interval of 30 seconds), the session terminated immediately. Ten trials were conducted, and the proposed system achieved a 100% success rate in terminating the session when the user's devices were not co-located.

### E. Computation overhead

The presented experiment measures the time required for the proposed authentication system to make decisions using six

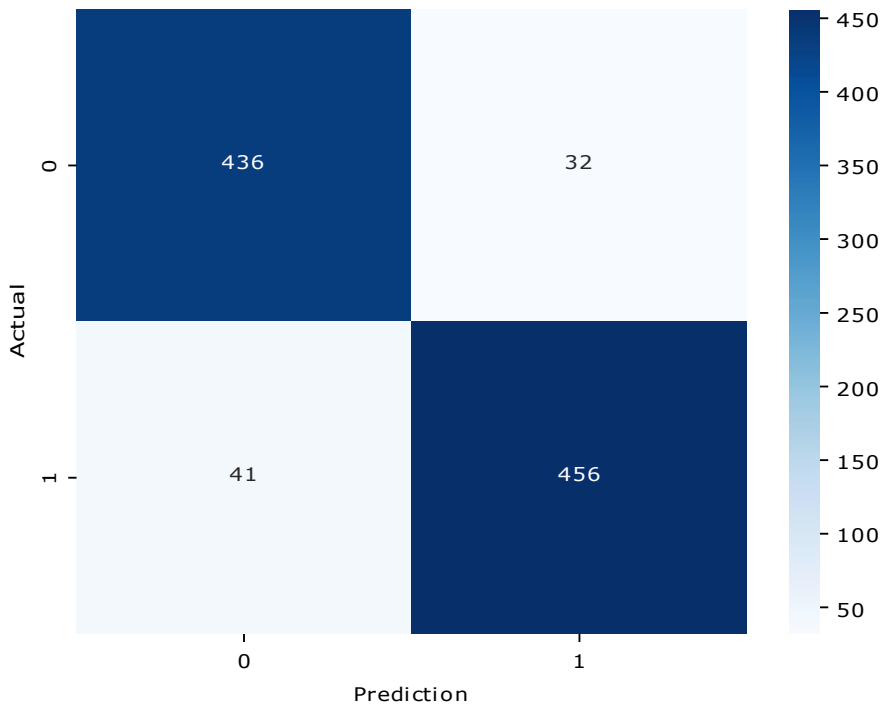

Decision Tree Confusion Matrix

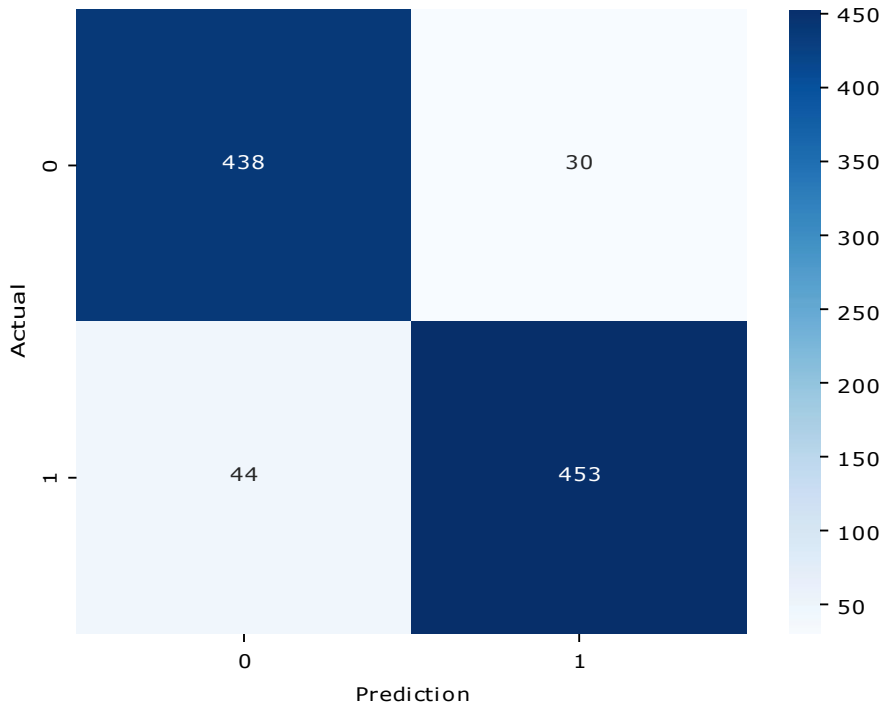

K-nearest Neighbors Confusion Matrix

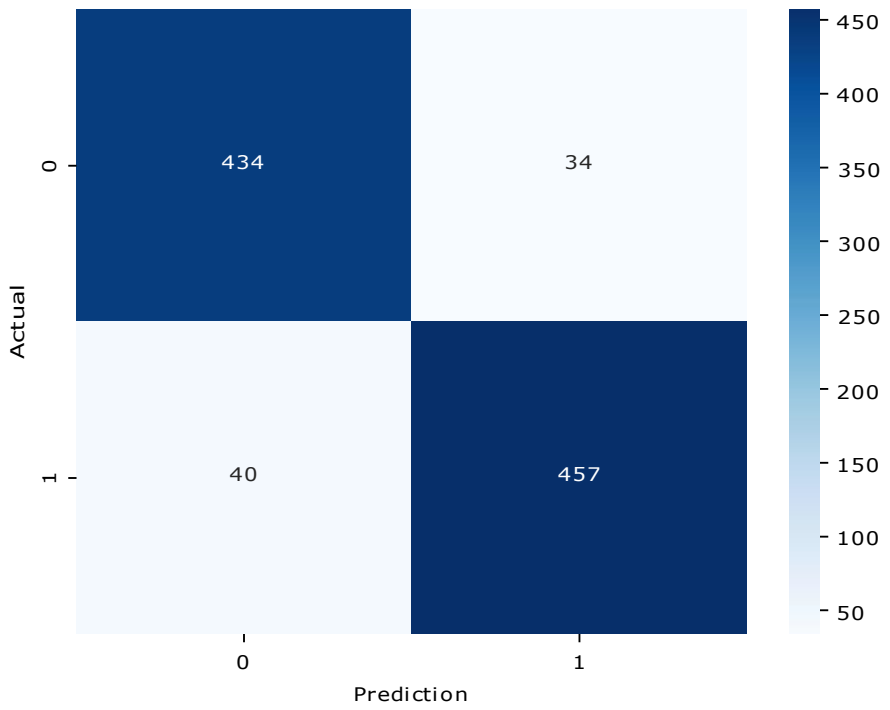

Random Forest Confusion Matrix

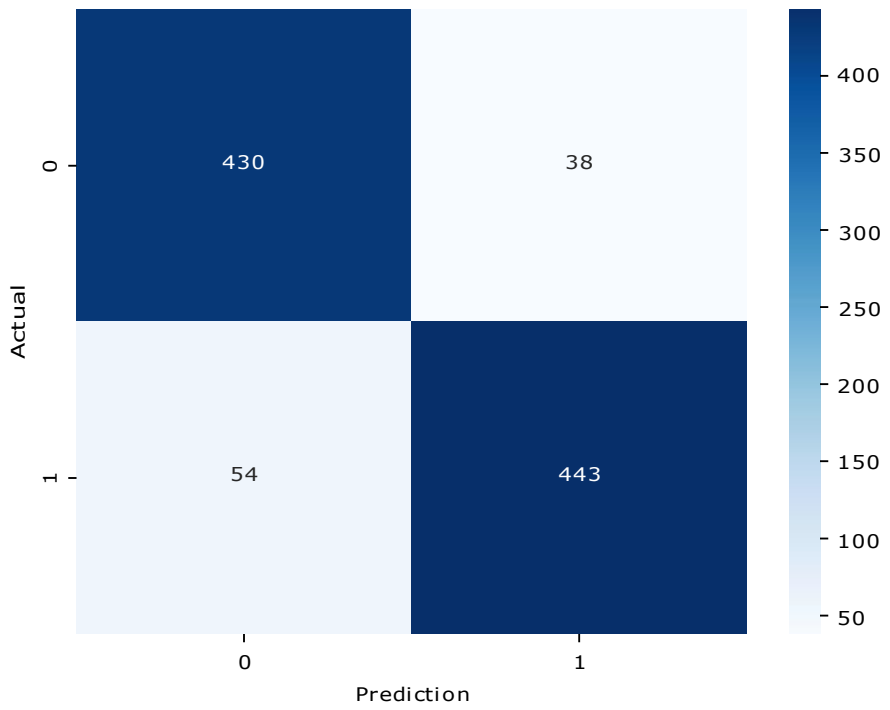

Support Vector Machine Confusion Matrix

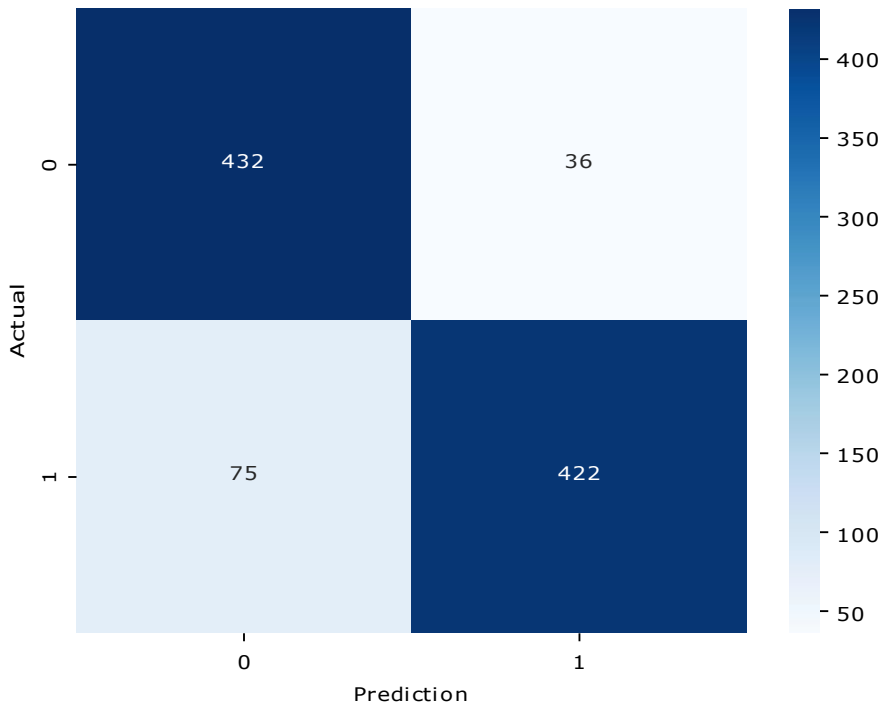

Naive Bayes Confusion Matrix

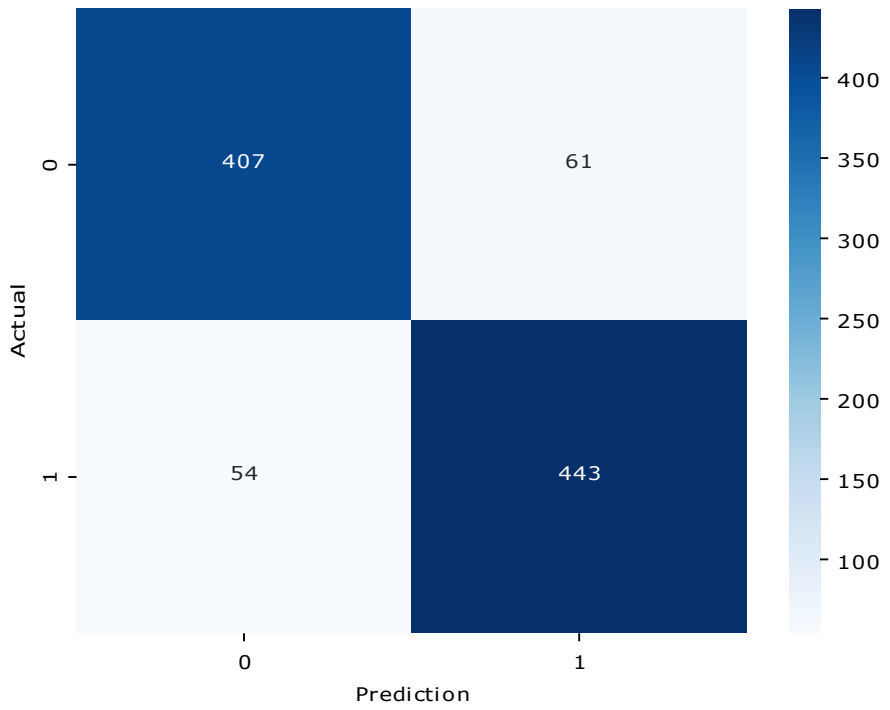

Logistic Regression Confusion Matrix

Fig. 4: Confusion matrices for various classifiers

different ML models: DT, KNN, RF, SVM, NB, and LR. The processing time for each ML model is reported in seconds in Table IV.

Based on the results, the proposed authentication system demonstrates efficiency in implementing the second layer of authentication, with the maximum time taken by any ML model being 0.263 s, indicating satisfactory overhead for the processing time.

TABLE IV: Time Consumption

| Model | DT | KNN | RF | SVM | NB | LR |
|---|---|---|---|---|---|---|
| Time | 0.002 s | 0.009 s | 0.1 s | 0.263 s | 0.001 s | 0.007 s |

### *F. Quantitative Comparison With The Discussed Related Works*

The quantitative comparative analysis presented in Table V underscores a notable trend in the extant literature: a predominant focus on accuracy, often at the expense of addressing efficiency. While many studies have achieved commendable accuracy rates, the conspicuous absence of efficiency metrics suggests a potential oversight in holistic model evaluation.

Our research endeavors to bridge this gap. By not only achieving a competitive accuracy rate of 92.4% but also emphasizing the model's efficiency, clocking in between 0.001s to 0.263s, we present a more comprehensive evaluation framework. This balanced approach underscores our model's applicability in diverse real-world scenarios, catering to both the need for correct classifications and timely outputs. Such a dual-focused evaluation approach is pivotal in advancing the field, ensuring that future models are not only accurate but also pragmatically efficient.

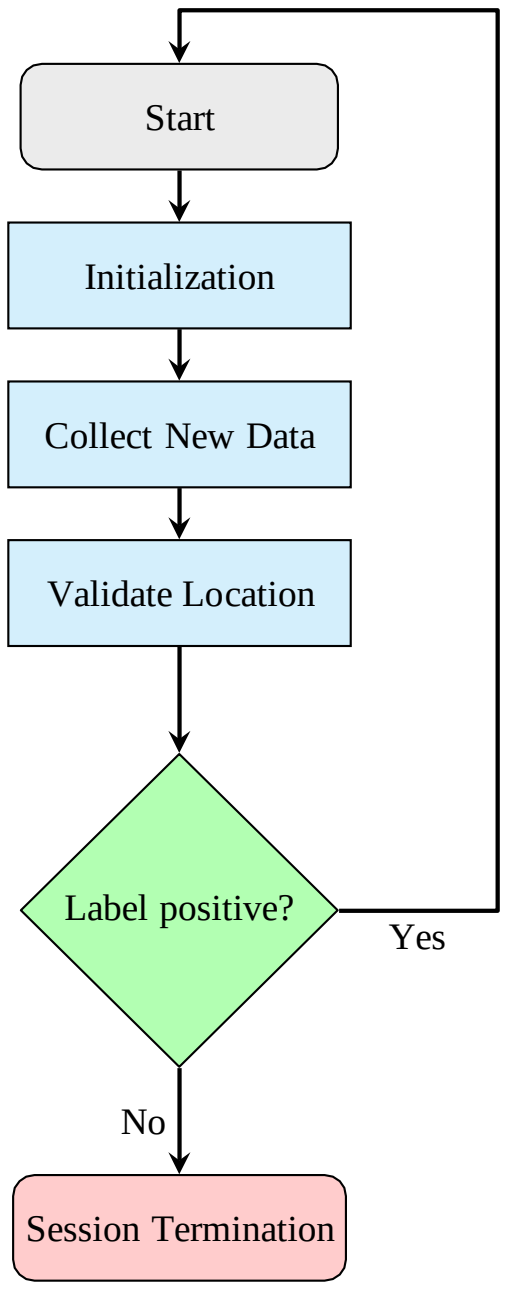

Fig. 5: Continuous Authentication Fluxogram

One of the standout features of our research is the deployment of a three-factor authentication system, highlighted in the "Number of Factor" column. While many conventional methods lean on 1-2 factors, typically passwords or biometrics, our approach presents a distinctive triad.

First, there is the Explicit Authentication, where users are prompted to input a valid username and password. Following that, we introduce two continuous, non-interactive, or "zero-effort" phases: Proximity and Device Presence and Zero-Effort Authentication. In the former, authentication is contingent upon both user devices (i.e., a login device and a mobile device) being present within overlapping Wi-Fi access points and staying within a designated proximity threshold. The latter employs ML to effortlessly authenticate users based on beacon frame characteristics and RSSI values from their devices, without requiring any further interaction from them.

TABLE V: Quantitative Comparison with The Discussed Works, where NM: Not Mentioned

| Study | Number of Factor | Accuracy | Time Consumption |
|---|---|---|---|
| [10]–[24] | 1-2 | 87% - 98.5% | NM |
| [25]–[38] | 1-2 | 90% - 99.11% | NM |
| Ours | 3 | 92.4% | 0.001s - 0.263s |

## VI. Security Analysis

In this section, we examine the security of the proposed 2FA system and assess its resilience against various cyberattacks.

### A. Evasion Attacks

In the context of the proposed 2FA system, an evasion attack could occur if an attacker attempts to manipulate the RSSI values to deceive the system. Continuous authentication is a key feature of the proposed system that can mitigate evasion attacks. By continuously monitoring the user's environment, the system can detect anomalies in the RSSI values that may indicate an attack attempt. The system can then take appropriate actions, such as requesting additional authentication from the user, or even blocking access to the account.

To evaluate the system's resilience against such attacks, we assess the system's performance when an attacker manipulates the RSSI values to deceive the system. We simulate an evasion attack by adding random noise to the RSSI feature of the test data and evaluating the system's performance on the modified test data. To this end, we fit each ML model to the training data and then test the system's performance on the noisy test data, see Figure 6. As shown in Figure 9a, the system's performance remains relatively stable for most of the classifiers even when random noise is added to the RSSI feature of the test data, indicating that the system is not significantly affected by an attack attempt. This result suggests that the proposed system is robust against evasion attacks.

### B. Model Extraction Attack

Model Extraction attack involves an adversary attempting to extract the parameters or structure of a ML model by training a separate "shadow model" based on the outputs of the original model. Once the shadow model is trained, the adversary can use it to make predictions about new data and compare these predictions to the outputs of the original model to learn more about the structure and parameters of the original model. To mitigate the Model Extraction attack, we use model ensembling, which involves training multiple models on the same dataset and combining their outputs to make a prediction. This makes it more challenging for attackers to extract the parameters of any single model, since they would need to extract the parameters of all of the models to gain a complete understanding of the system.

We loaded a dataset and trained a RandomForest (RF) model on it. To simulate a Model Extraction attack, we created adversarial examples by significantly altering the RSSI (dBm) values in the dataset. Using these adversarial examples and the predictions from our original RF model, we trained a shadow RF model. We then evaluated the shadow model's performance on the original test data to gauge its ability to approximate the original RF model's behavior. This entire process is detailed in Figure 7.

Figure 9b shows the performance of the attack model, indicating that the targeted model (RF) was affected by this type of attack, while the rest were not affected. These results demonstrate the effectiveness of our proposed system's model ensembling approach in mitigating the Model Extraction attack.

### C. Radio Frequency Signal Interference Attack

Radio Frequency Signal Interference Attack is a cyberattack that disrupts wireless communication between two devices

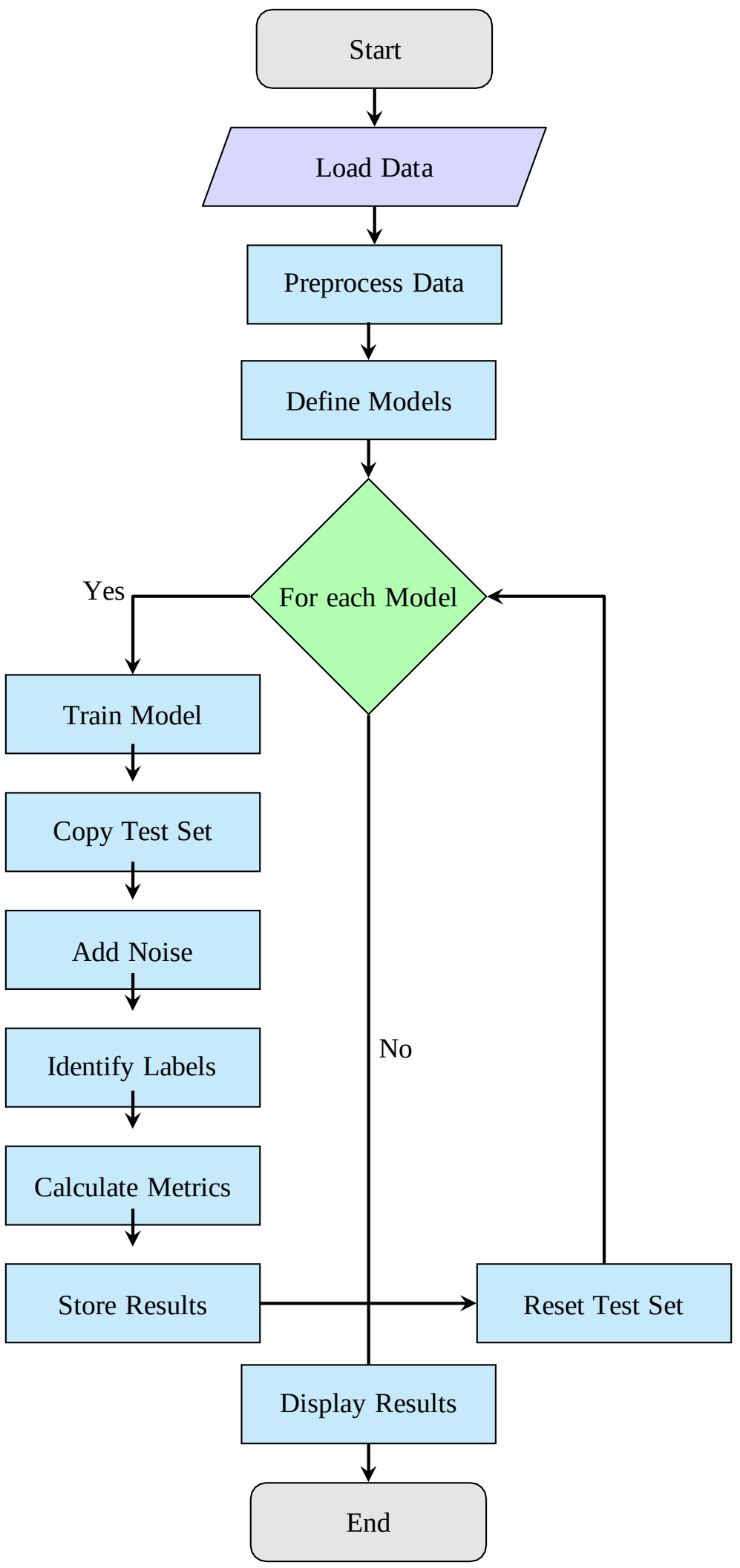

Fig. 6: Evasion Attack Simulation Fluxogram

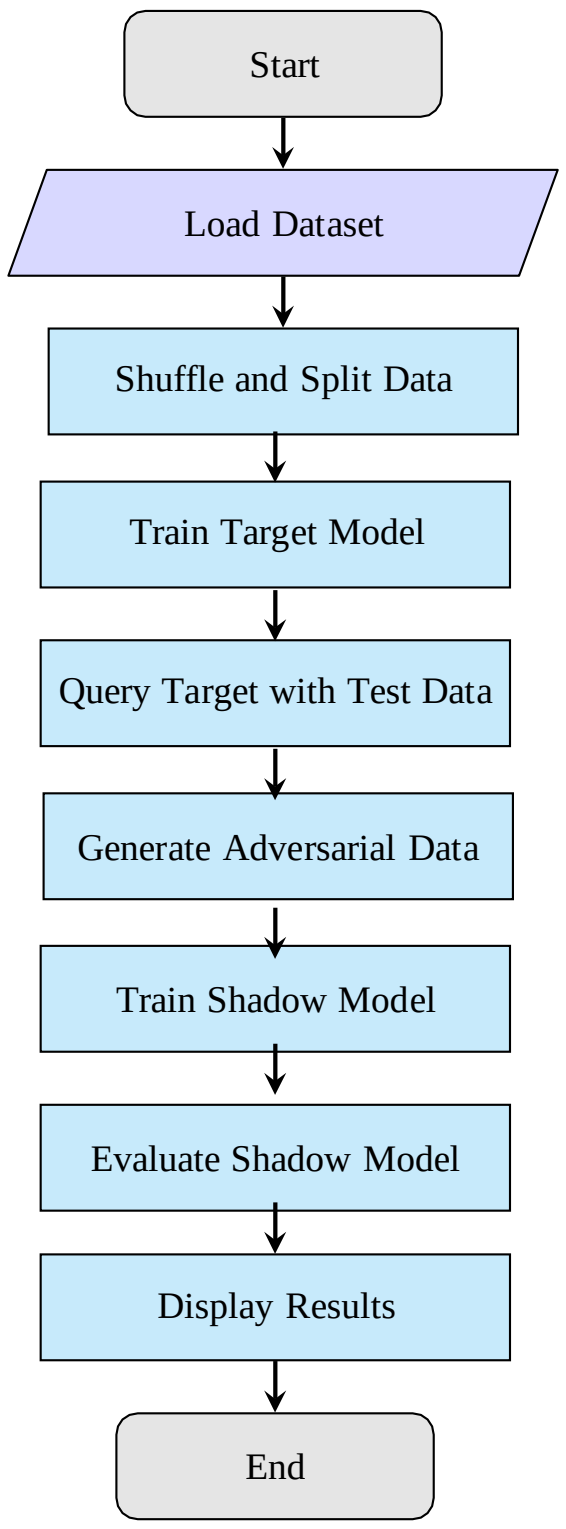

Fig. 7: Model Extraction Attack Simulation Fluxogram

(i.e., a login device and a mobile device) by transmitting radio signals. The attacker injects unauthorized radio signals into the wireless network, causing signal degradation, packet loss, and potentially unauthorized access to sensitive information. Continuous authentication can also help mitigate the effects of this attack by constantly monitoring the user's environment and comparing it with the data collected during the initial login. If there is a significant deviation from the original data, the system can assume that an attack is underway and either deny access or request additional authentication measures.

We simulated RF signal interference attack by adding random noise to the feature data using NumPy's *'random.normal()'* function; see Figure 8. The Interference variable is generated using this function with a normal distribution having a mean of zero and a standard deviation of 2. The noise is added to the feature data *'x'* by adding Interference to it, creating a new array *'x_noisy'*.

The results of our experiment indicate that the proposed system demonstrated a high level of resilience to this type of attack, even when subjected to significant interference. Figure 9c provides a visual representation of the results obtained.

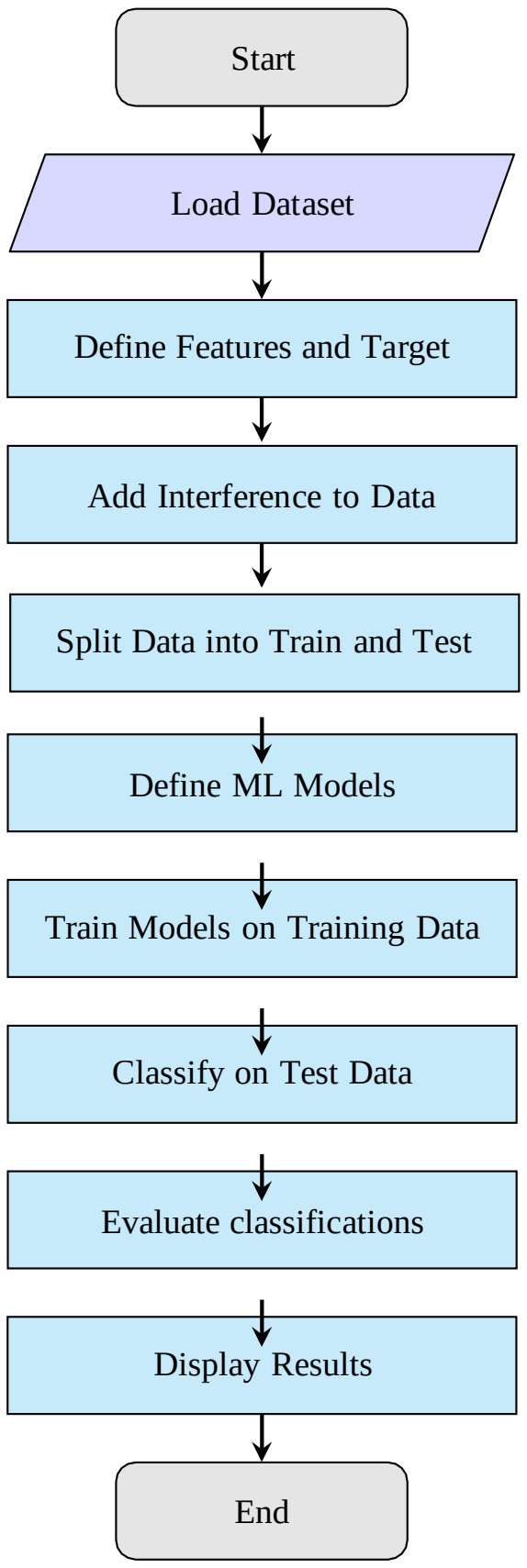

Fig. 8: Radio Frequency Signal Interference Attack Simulation Fluxogram

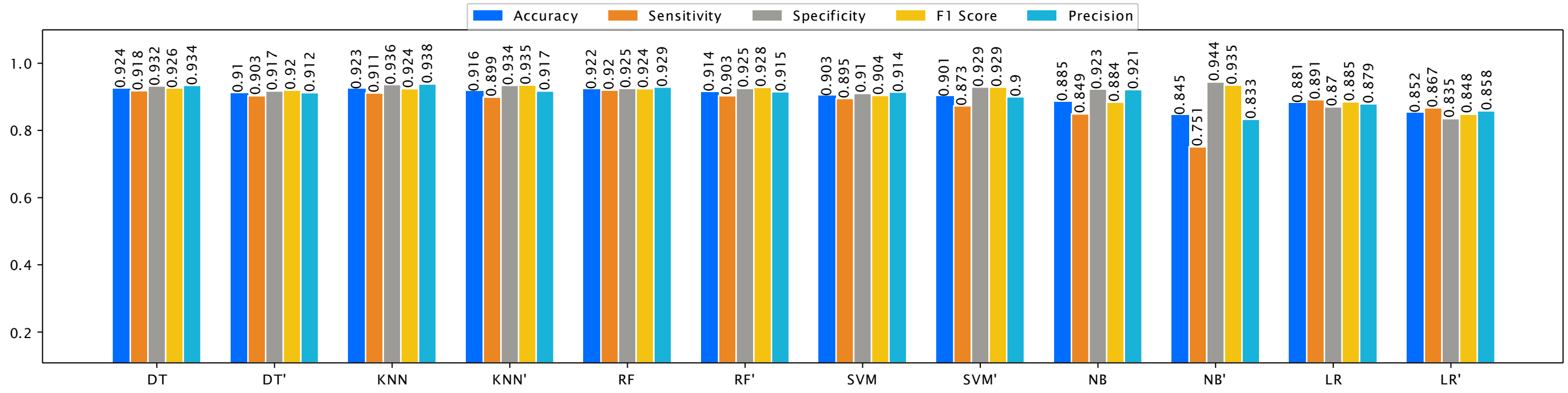

(a) Simulate Evasion Attack Result Visualization

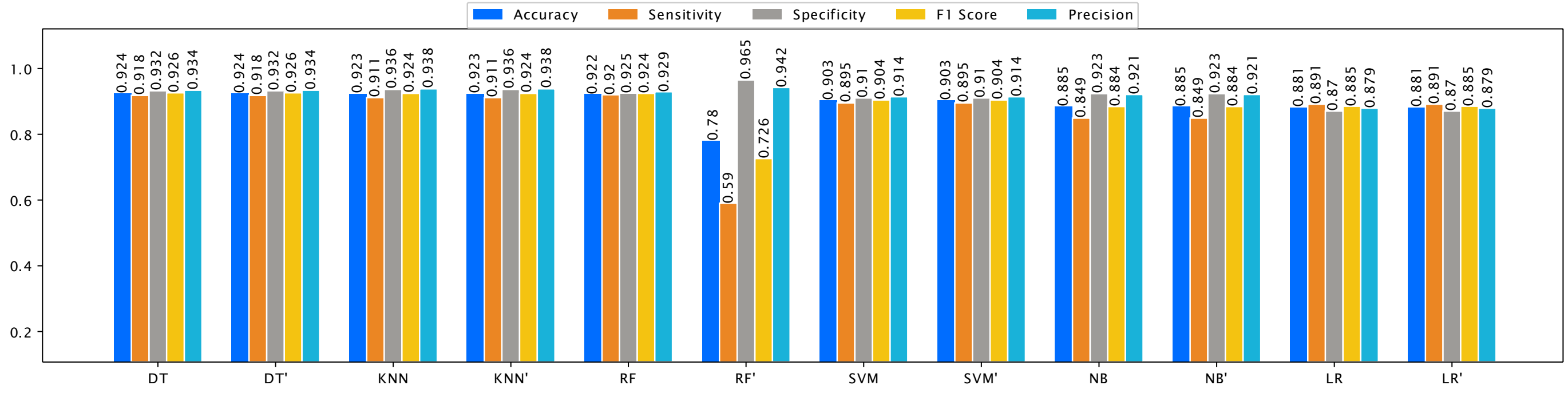

(b) Simulate Model Extraction Attack Result Visualization

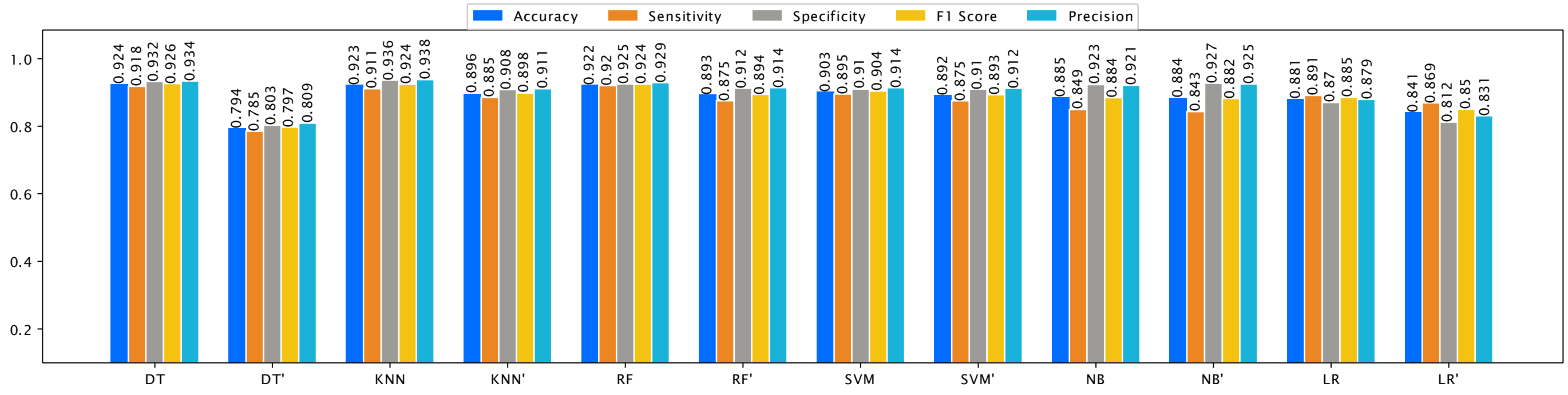

(c) Simulate RF Signal Interference Attack Result Visualization

Fig. 9: Combined Attack Result Visualizations; Modal's Name with Apostrophe Indicates Result After Attack

## VII. System Analysis

In this section, we discuss the proposed system features, each one in its own subsection.

### A. Reliability

ML is employed in our authentication system to bolster reliability and precision, circumventing the limitations of traditional two-factor methods that often necessitate manual input of codes or tokens. By analyzing the user's environment, ML eradicates the need for user input and mitigates human errors, thereby enhancing the system's dependability. The self-adaptive capability of ML minimizes the necessity for frequent maintenance or updates, rendering the system cost-efficient.

Our proposed scheme demonstrates outstanding reliability, as evidenced by a 92.4% success rate in implementing the second layer of authentication in our experimental results, as presented in Table II. Before implementing the second layer of authentication, the system must achieve a 100% success rate in authentication steps number 1, 2, and 4 in Figure 2. We further evaluated its robustness under adversarial conditions by introducing various attack scenarios in Section VI. Even in these challenging circumstances, the highest time for a successful login using ML remained at a swift 0.263 seconds as can be seen from Table IV. Collectively, these factors contribute to the system's robust performance, instilling confidence in its practical, real-world applications.

### B. Scalability

The proposed system that utilizes broadcast messages from Wi-Fi APs to authenticate users relies heavily on the scalability of the system to handle an increasing number of users and devices smoothly. Smart devices, like mobile phones, laptops, and tablets, can detect signals from various radio frequency sources in their vicinity, and as they move between networks, they receive broadcast messages from APs containing useful information.

This allows for high scalability, as any Wi-Fi-enabled device within range can receive and process the messages, regardless of its connection to a particular network (i.e., used in the authentication process, step number 3 in Figure 2). The ability of the proposed system to scale effortlessly enables it to handle a large number of users and devices simultaneously without any significant impact on the system's performance. This feature makes the proposed system an excellent choice for organizations that require a scalable security system to manage their growing user base without compromising the system's reliability or performance.

### C. Limitation and Contingency

In situations where a user cannot access his/her mobile device or there are no APs in their vicinity, the proposed system offers a one-time login solution. This solution requires the user to provide his/her username and answer a security question. Upon successful completion, a one-time Password (OTP) is sent to the user's registered email address, which they can use to access resources granted through the authentication entity. Although this solution goes against the proposed system's fundamental requirement of not requiring user interaction, it is intended to be used only in rare cases when the user does not have access to his/her mobile device. In typical settings, this should not be a common occurrence. Hence, the one-time login solution serves as a backup mechanism for cases where the user cannot access his/her mobile device.

The one-time login solution ensures that users can access resources even when they are not near an AP or do not have access to his/her mobile device. As a result, it enhances the user experience by reducing potential downtime and frustration caused by technical limitations.

## VIII. Conclusion

This paper introduces a Two-factor Authentication (2FA) system that leverages radio waves and Machine Learning (ML) to create an efficient and reliable user authentication process. The proposed system can adapt to different environments and locations and can be customized to meet the specific security needs of a protected entity. The system's parameters, such as the number of Wi-Fi Access Points (APs) and distance threshold, can be adjusted to fit the desired level of security. The proposed system does not require any specialized hardware or infrastructure, making it cost-effective and easy to maintain. The system also uses a ”zero-effort” approach, automating the authentication process and reducing the risk of human error while enhancing user experience and security. Moreover, the continuous authentication feature verifies the user's identity constantly and terminates the session if the devices (i.e., a login device and a mobile device) are no longer in the same location, further improving security and user convenience.

Future research might include a thorough evaluation of the system's performance under various practical scenarios, such as public places, hospitals, offices, and government buildings, to better understand the system's capabilities and limitations. Investigation into how to integrate the proposed system with existing systems, such as access control systems, and how to make the system more scalable and efficient would also be useful for practical implementation and cost optimization.